# Quantum electromechanics: Qubits from buckling nanobars

Sergey Savel'ev[(1)], Xuedong Hu[(1,2)], and Franco Nori[(1,3)]

[(1)] *Frontier Research System, The Institute of Physical and Chemical Research (RIKEN), Wako-shi, Saitama, 351-0198, Japan*
[(2)] *Department of Physics, University at Buffalo, SUNY, Buffalo, NY 14260-1500, USA and*
[(3)] *Center for Theoretical Physics, Department of Physics,*
*University of Michigan, Ann Arbor, MI 48109-1120, USA*
(Dated: October 19, 2004)

We propose a mechanical qubit based on buckling nanobars—a NEMS so small as to be quantum coherent. To establish buckling nanobars as legitimate candidates for qubits, we calculate the effective buckling potential that produces the two-level system and identify the tunnel coupling between the two local states. We propose different designs of nanomechanical qubits and describe how they can be manipulated. Also we outline possible decoherence channels and detection schemes. A comparison between the well studied charge and flux superconducting qubits suggests several experimental setups which could be realized using available technology.

Micro- and nano-electromechanical systems (MEMS and NEMS) have attracted widespread attention because of their broad spectrum of functionalities, their tiny sub-micron sizes, and their unique position bridging microelectronic and mechanical functions[1]. Sophisticated tools ranging from mirrors and sensors, to motors and multi-functional devices have been fabricated[2]. As the size of these devices shrinks, experimental studies of NEMs are rapidly approaching the quantum limit of mechanical oscillations[3–5], where quantum coherence and superposition should result in quantum parallelism and the possibility of information processing. Many microscopic objects, such as electron and nuclear spins, have been suggested as quantum bits (qubits), the building block of a quantum computer.

The emergence of quantum electromechanical devices (see, e.g., Ref.[1–5] and references therein) brings both challenges, such as the inevitable and ubiquitous quantum noise, and promises, such as macroscopic quantum coherence[5–7] or quantum teleportation[8,9]. Indeed, during the past several years the quantum mechanical properties of NEMS and how they can be coupled to other quantum mechanical objects have been very actively studied[10–15]. The experimental pursuit of these studies has so far focused on cooling a device to reach states where quantum fluctuations in the lowest energy eigenmode dominate over thermal fluctuations[3,11]. Such eigenmodes are generally harmonic-oscillator modes with equal energy-spacings and follow bosonic statistics.

Among NEMS there also exist systems that can be well approximated by two levels in the quantum limit, so that they might be candidates for qubits. For example, when a longitudinal strain above a certain critical value is applied to a small bar with one or two ends fixed, there exist two degenerate buckling modes, as schematically shown in Fig. 1. In the quantum mechanical limit, these two modes represent the two lowest-energy states. They can be well separated from the higher-energy excited states, so that at low temperatures the buckling nanobar can be properly described as a two-level system. Furthermore, since nanobars can be charged, the electric field can be conveniently used to manipulate their quantum states.

The fascinating prospect of observing quantum coherent phenomena in a macroscopic mechanical oscillator is a main motivation of this study. To achieve this goal, a NEMS needs to possess two general attributes: small size and a high fundamental frequency. A carbon nanotube is a natural candidate: it is very thin (and can be very short) while still being stiff[16] (thus having high vibration frequency) because of the strong covalent bonds between carbon atoms within the graphene sheet[17,18]. Recent experiments[19] on suspended charged carbon nanotubes excited by a time-periodic electric field show a pronounced anharmonic behavior. Moreover, the resonance peak for one of the fundamental harmonics first increases, as expected, and then surprisingly splits into two sub-peaks when increasing the amplitude of the externally applied electric field[19]. This unusual behavior proves[19,20] that the *suspended carbon nanotube* can be described as a particle (associated with the center of mass of the nanotube) moving in a *double-well potential*[19,20]. This nanotube jumps between the two potential wells due to either thermal activation or quantum tunnelling, depending on temperature. Thus, the technology for fabricating the suspended buckled carbon nanotubes, a candidate for a nanomechanical qubit, already exists[19]. In view of the explosive growth of NEMS technology, here we discuss the prospect of such buckling charged nanobars (the clamping at the base ensures an anisotropic nanobar instead of an isotropic nanotube) as candidates of quantum bits for quantum information processing.

*Quantum state in a double-well potential.*— The double-well potential corresponding to the two buckling modes takes the form: $U(y) = -\alpha y^2 + \beta y^4$, with $\alpha > 0$, $\beta > 0$ (Fig. 1a). This potential has two minima at $y = \pm y_0 = \pm\sqrt{\alpha/2\beta}$ that are separated by a potential barrier $\Delta U = \frac{\alpha^2}{4\beta}$. The first two energy levels $E_1$ and $E_2$ in the right well can be estimated assuming a parabolic potential well shape $U \approx m\omega^2(y-y_0)^2/2$ with $\omega = \sqrt{U''(y_0)/m} = 2\sqrt{\alpha/m}$. Thus we obtain $E_1 = 3\hbar\sqrt{\alpha/m}$ and $E_2 = 5E_1/3$. Due to the quantum tunnelling between left and right potential wells, $E_1$ splits into two levels $E_1^\pm = E_1 \pm \hbar\Delta_t$. The tunnelling rate $\Delta_t$ between the left and right buckled states is:[21]

$$\Delta_t \approx \frac{2}{\pi}\sqrt{\frac{\alpha}{m}} \exp\left(-\frac{\pi\sqrt{2m}(\Delta U - E_1)}{2\hbar\sqrt{\alpha}}\right), \quad (1)$$

where $\omega_0$ is the classical oscillation frequency in each well, $a$ is the location of the turning point, corresponding to the energy $-\Delta U + E_1$.

Next we compare the quantum tunnelling probability $P_q$ and the thermal activation probability $P_T$ of transitions between the two buckled states $|R\rangle$ and $|L\rangle$, which would establish the threshold temperature below which quantum behavior can be observed. These probabilities could be simply estimated as $P_q \propto \exp\left(-(2/\hbar)\int_{-a}^{a}|p|dx\right)$ and $P_T \propto \exp\left(-(\Delta U - E_1)/k_B T\right)$, where $T$ is the temperature and $k_B$ is the Boltzmann constant. Thus, the conditions for quantum tunnelling to dominate thermal activation becomes: $T < T_{\text{crossover}} = \hbar\alpha/(k_B\pi\sqrt{2m})$. Below we estimate the crossover temperature for a compressed charged elastic rod in a transverse electric field, which has drawn a lot of theoretical interest (see, e.g., Ref.[5]).

The energy of a compressed charged rod in an external electric field can be written as[22]:

$$F_b = \int_0^{l_{\max}} dl \left\{ \frac{IE(y'')^2}{2Y^2(l)} + f(Y(l)-1) + yf_\perp \right\}. \quad (2)$$

where $y(l)$ is the transverse deviation from the straight position, parameterized by the arclength $l$ ($0 \leq l \leq l_{\max}$) and $Y(l) = \sqrt{1-(y'(l))^2}$. Here we introduce the elastic modulus $E$ and the moment of inertia $I$ of the rod, the mechanical force $f$ acting on the end of the rod in the longitudinal direction, and the transverse force $f_\perp$. Hereafter, we use the notation $d/dl \equiv '$. If the rod is charged, the transverse electric field $E_\perp$ provided by a capacitor (Fig. 1b) can be used to precisely control the transverse force $f_\perp$. Note that even a rather small charge $Q$, which does not affect the stiffness of the rod, is enough to achieve a desirable value of $f_\perp$, given a sufficiently strong electric field. We assume a buckling mode $y(l) = y_0 \sin(\pi l/l_{\max})$, which corresponds to a rod with hinged ends. The particular choice of the boundary conditions at $y(0)$ and $y(l_{\max})$ (e.g., $y(0) = y'(0) = y(l_{\max}) = y'(l_{\max}) = 0$) does not affect the results. Substituting $y(l)$ into Eq. (5), and expanding $F_b$ up to $y_0^4$, we obtain the potential energy $U$ as a function of the collective buckling coordinate $y_0$: $U(y_0) = \alpha_1 y_0^2 + \beta_1 y_0^4 + (2l_{\max}f_\perp/\pi)y_0$, where[23]

$$\alpha_1 = \frac{\pi^2}{4l_{\max}}(f_c - f), \quad \beta_1 = \frac{\pi^4}{64l_{\max}^3}\left(\frac{4IE\pi^2}{l_{\max}^2} - 3f\right), \quad (3)$$

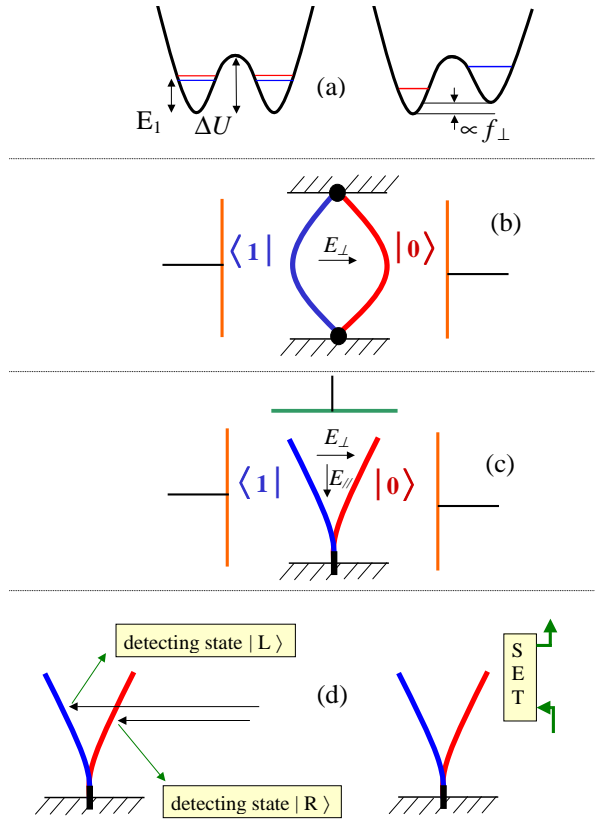

FIG. 1: (a) Double-well potential for a buckled nanobar. Due to tunnelling from the right potential well to the left one, the lowest energy level is split into two levels for $f_\perp = 0$ as shown in the left panel. The lowest (blue) and the first excited (red) levels correspond to the symmetric and antisymmetric combinations of the wave functions localized in the left and right potential wells. The energy level splitting between the left and right states can be controlled by the transverse force $f_\perp$ as shown in the top right panel. (b) A buckled rod qubit where the compressed force applied to the rod ends controls the potential shape [$\alpha_1$ and $\beta_1$ in Eq. (3)] and, therefore, the energy splitting at the degeneracy point. The transverse force $f_\perp$ allows to drive the bar to a degeneracy point. (c) Another proposed design for a buckled-rod qubit. In this case the possible decoherence originating from the relative vibration of the top and bottom rod holders can be avoided. By changing the charge of the top capacitor plate and the charge of the rod itself it is possible to independently control both the parallel-to-rod electric field $E_\parallel$ and its gradient. This design allows a high level of control of both the tunnelling and the energy splitting. (d) Single-shot measurements could be done via either single electron transistor (SET) or photon reflection.

where $f_c = IE\pi^2/l_{\max}^2$. This leads to

$$T_{\text{crossover}} \gtrsim \left(\frac{3\hbar}{\sqrt{m}}\frac{\pi^6 IE}{16\, l_{\max}^5}\right)^{1/3} \frac{\hbar}{\pi k_B \sqrt{2m}} \quad (4)$$

We obtain $T_{\text{crossover}} \gtrsim 1.35\, K/l_{\max}^{2.33}$, with $l_{\max}$ measured in nanometer.

*Elastic rod in longitudinal and transverse electric fields.*— We now consider a novel and far more controllable design of a nanorod qubit, which is also free from decoherence coming from the relative vibrations of the rod ends (Fig. 1c). In such a design we consider a "crossed" electric field having a component $E_\parallel$ along the rod and another $E_\perp$ perpendicular to the rod, which is clamped to a substrate only by one end. In principle, changing the charge of the top capacitor plate and the charge of the rod itself can control both the electric field, $E_\parallel$, on the rod and its gradient, $\partial E_\parallel / \partial x$. The energy of the charged rod can be written as

$$F_b = \int_0^{l_{\max}} dl \left\{ \frac{IE(y''(l))^2}{2Y^2(l)} + f_\parallel \int_0^l ds \, (Y(s)-1) \right.$$
$$\left. + f'_\parallel \left[ \int_0^l ds \, (Y(s)-1) \right]^2 + y(l) \cdot f_\perp \right\}. \quad (5)$$

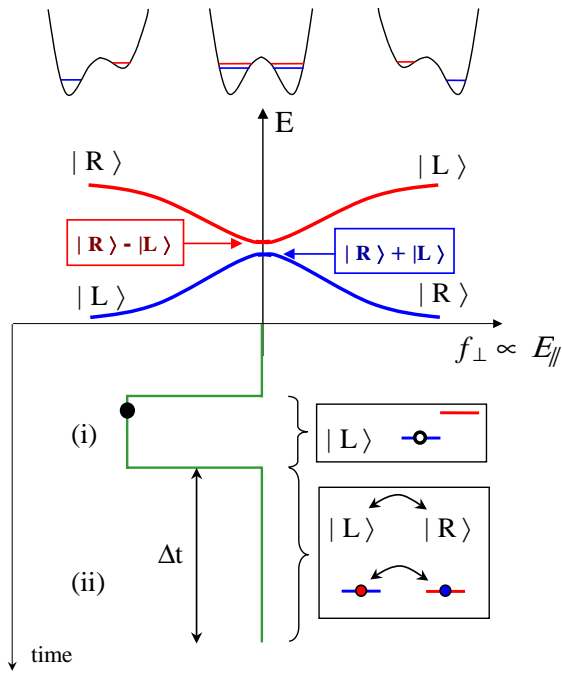

FIG. 2: A schematic diagram for controlling the quantum state of a nanobar via coherent oscillations at the degeneracy point. Tuning the system to its degeneracy point can be done by changing the perpendicular electric field $E_\perp$. For the case when $E_\perp$ and $E_\parallel$ is zero at $t=0$, the perpendicular electrical field $E_\perp$ has to be applied first and then the longitudinal field $E_\parallel$ (e.g., at the moment marked by the solid circle on the $E_\perp$(time) curve). This initializes the system during the stage (i). Switching off the perpendicular electric field $E_\perp$ brings the system back to its degeneracy point, and the nanorod starts to oscillate a time $\Delta t$ during stage (ii). To observe coherent oscillations, measurements (e.g., optically or electrically) must be done for many values of $\Delta t$.

By adjusting the different knobs, we can tune the relative strength between thermal activation and quantum tunneling, allowing the observation of transition between these two regimes.

*Manipulation, decoherence, and detection of mechanical qubits.*— The manipulation of the mechanical qubits can be achieved electrically (Fig. 2). For example, in analogy to the Cooper pair box[24] (see table), one can prepare the nanobar qubit in the $|L\rangle$ state by setting a transverse electric field towards the right (assuming the nanobar is negatively charged). By *suddenly* turning off this electric field and bringing the system to the degeneracy point, the nanobar state is prepared in a coherent superposition of $(|R\rangle \pm |L\rangle)/\sqrt{2}$. Because of the tunnel splitting, the system then starts to oscillate coherently, with a frequency given by $\Delta_t$. After a period of time $\Delta t$, the qubit can be in either the $|L\rangle$ or $|R\rangle$ states. Therefore, by detecting the nanobar position, as a function of $\Delta t$ (see Fig. 2), one can determine the coherent oscillation frequency and the system decoherence. Driven coherent transitions between the two qubit states can also be similarly achieved. A sinusoidal component can be added to the vertical electric field that is used to control the tunnel coupling between the $|L\rangle$ and $|R\rangle$ states, in analogy to the microwave driving force on the Josephson coupling in a Cooper pair box. The study of free and driven coherent oscillations of a nanobar would help clarify its quantum coherence properties and demonstrate its feasibility as a qubit.

| System | SC charge QB | SC flux QB | Nano-bar QB |
|---|---|---|---|
| States | \| excess charge $>$ : $\|0\rangle$ and $\|1\rangle$ | \| current direction $>$ : $\|\circlearrowleft\rangle$ and $\|\circlearrowright\rangle$ | \| buckling direction $>$ : $\|R\rangle$ and $\|L\rangle$ |
| Hamiltonian | $H = \epsilon \, \sigma_z + \hbar \, \Delta_t \, \sigma_x$ | | |
| Tunneling $\Delta_t$ and energy splitting controlled by | Gate voltage $V_g$ (normalized: $n_g$) | Magnetic flux: $\Phi$ (normalized: $f = \Phi/\Phi_0$) | Transverse force $f_\perp$ (induced, e.g., by a transverse electric field $E_\perp$) |
| Tunneling controlled by | Josephson energy: $\mathcal{E}_J(\Phi)$ | Josephson energy: $\mathcal{E}_J$ | Either • **Longitudinal force** (induced by pressure or electrical field $f_\parallel$) • or its **gradient** $E_\parallel$ Longitudinal force and its gradient control $\partial_x f_\parallel$ barrier height and its curvature |
| Coupling between qubits | **Electrical** (e.g., capacitive or inductive coupling) | **Magnetic** | **Electrical** (dipolar) |
| Decoherence sources include | **Charge** fluctuations | **Flux** fluctuations | 1. **Charge** fluctuations 2. Phonon-phonon interactions |
| Read-out | **Electrical** (e.g., SET or JJ) | **Magnetic** (SQUID) | Either **electric** or **optical**, or **mechanical** |

Table: comparison of Josephson-junction superconducting (JJ SC) charge, JJ SC flux and nano-bar qubits (QB).

Universal quantum computing requires two-qubit operations. For charged nanobars, the inter-qubit interaction comes naturally in terms of the electric dipole interaction between the bars, quite similar to the dipole interaction used in other proposed qubits. In the case of nanobar qubits, one can again use the transverse electric field $E_T$ to tune selected qubits into resonance, then apply microwaves to perform conditional rotations and other operations.

For NEMS, and for more conventional applications such as resonators, major sources of noise[25,26] (that limit the quality factor $Q$) are internal thermomechanical noise (such as heat flows due to inhomogeneous distribution of strain), Nyquist-Johnson noise from the driving electrical circuit, adsorption-desorption noise when a resonantor is moving in a non-vacuum, noise from moving defects, etc. In the quantum mechanical limit, some of these noises become unimportant. For example, there should be no heat flow: the nanobar should not be driven beyond its first excited state. The NEMS can be placed in a vacuum to reduce adsorption-desorption noise. The whole system, including its electrical components, has to be cooled in a dilution refrigerator in order to reach the two lowest states of the nanobar, so that the Nyquist-Johnson noise is also suppressed. The main source of quantum mechanical decoherence might be internal dissipation caused by phonon-phonon interactions. Since the nanobars are clamped onto a much larger substrate, in which lower energy phonon modes exist, coupling at the base of the nanobars could lead to relaxation/excitation of the qubit states. Another source of decoherence would be charge fluctuations. Charged nanobars allow increased maneuverability of the NEMS; but the surrounding and internal charge fluctuators will lead to charge noises that cause relaxation and dephasing of the qubit.

Single-Wall Nano Tubes (SWNTs) are a natural candidate for the mechanical qubit we consider here. However, our proposal is not limited to them. Indeed, a multiple-wall nanotube could provide a higher-frequency mechanical oscillator and more favorable condition for observing macroscopic quantum tunnelling and the coherent evolution of mechanical motion, assuming motion between walls can be effectively pinned. Silicon-based systems provide another enticing alternative, especially from the perspective of fabrication. While it might be difficult to clamp several SWNTs to make identical nanobars (with controlled inter-bar distance, same length, same buckling orientation, and same response to external stress), fabricating a lithographically-patterned silicon-based nanostructure would be much more reliable. SiC nanobars have shown higher stiffness[18] compared to Si. Unless nano-assembled nanotubes can be made with sufficiently high precision, materials that can be lithographically fabricated seem more promising candidates for a larger-scale mechanical quantum information processor. This would be an ironic turn of events, given that the first computers (by C. Baggage) where mechanical.

The detection of the nanobar state can be done either electrically[3,14,27] or optically[28,29]. It has been shown in recent experiments[3] that single electron transistors (SET) are very sensitive to small charge displacements. Optical detectors can also be used, where light scattering can detect the state of the bent nanorod (Fig. 1d).

Very recently, Ref.[30] fabricated suspended nanobars (driven by a 25 MHz current through an attached electrode) switching between two distinct states. These suspended nanobars have already been tested[30] as very fast and very low-power-consumption storage memory devices. Still, many challenges lie ahead on the road to practical quantum electromechanics. We hope that our proposal here stimulate more research towards the ultimate quantum limit of NEMS.

This work was supported in part by the NSA and ARDA, under AFOSR contract number F49620-02-1-0334; and also by the US NSF grant No. EIA-0130383.